\documentstyle[12pt,psfig]{article}
\newcommand{\bea}{\begin{eqnarray}}
\newcommand{\eea}{\end{eqnarray}}
\newcommand{\be}{\begin{equation}}
\newcommand{\ee}{\end{equation}}
\newcommand{\vs}[1]{\vspace{#1 mm}}

\newcommand{\dsl}{\pa \kern-0.5em /}

\newcommand{\pa}{\partial}

\begin{document}
\topmargin 0pt
\oddsidemargin 0mm

\begin{flushright}
hep-th/0304084\\
\end{flushright}

\vs{2}
\begin{center}
{\Large \bf  
Accelerating cosmologies from M/String theory compactifications}
\vs{10}

{\large Shibaji Roy\footnote{E-Mail: roy@theory.saha.ernet.in}}
\vspace{5mm}

{\em 
 Saha Institute of Nuclear Physics,
 1/AF Bidhannagar, Calcutta-700 064, India\\}
\end{center}

\vspace{1.5cm}
\centerline{{\bf{Abstract}}}
\vs{5}
\begin{small}
We point out that the solution of $(4+n)$-dimensional gravity coupled to
the dilaton and an $n$-form field strength can give rise to a flat
4-dimensional universe (with a scale factor) of the type proposed recently
under time dependent compactifications. The compact internal spaces could be
hyperbolic, flat or spherical and the solution is identical to the space-like
two brane or S2-brane. As has been shown previously for SM2 solution with 
a fixed field strength we show that for $n=7$ (where the dilaton is 
vanishing and with a general field strength), 6 
the corresponding SM2 and SD2 solutions can give accelerating 
cosmologies in Einstein frame for both hyperbolic and flat internal 
spaces, thereby
meeting the challenge of obtaining such a solution from M/String theory
compactifications.
\end{small}

\vfil\eject

Since the discovery of the astronomical observation \cite{perl,rie} 
supported by the recent
measurement \cite{lb,be} of cosmic microwave background that our 
universe is undergoing
an accelerated expansion, it has been a major 
challenge \cite{mn,tow,gibhul} to obtain such models from M/string 
theory compactifications.
Recently Townsend and Wohlfarth (TW) \cite{tw} 
have shown in a beautiful paper that such
accelerated cosmologies can arise from a solution of $(4+n)$-dimensional
vacuum Einstein equation compactified on $n$-dimensional hyperbolic space
of time varying volume. For $n=7$ this implies that realistic cosmology can
result from M-theory compactification on a hyperbolic space. Interestingly,
although the original theory does not violate the strong energy condition,
such compactification does violate this condition
and accelerated expansion is obtained in Einstein frame circumventing the
``no-go'' theorem \cite{mn} which forbids acceleration in standard 
compactification.

In this paper we show that not only the solution of vacuum Einstein equation,
but also that of the $(4+n)$-dimensional gravity coupled to the dilaton and
an $n$-form field strength gives rise to the flat 4-dimensional universe
(with a scale factor) of the type discussed by TW. The solution of the above 
system has already been obtained \cite{cgg} under the name space-like 
branes or S-branes \cite{gs}\footnote{See \cite{kmp,roy,dk}
for some 
related works on S-branes.}.
S$p$-branes are time-dependent solutions with $(p+1)$-dimensional Euclidean
world-volume and apart from time they have $(d-p-2)$-dimensional hyperbolic,
flat or spherical spaces as transverse spaces. They can be understood to arise
as a time-like tachyonic kink solution of non-BPS D$(p+1)$ branes \cite{sen}
in string 
theory and might be useful to understand the time-like holography of the 
dS/CFT correspondence \cite{str}. It is clear from above that in 
order to get a 
4-dimensional world $p$ should be 2 and so we will look into the S2-brane
solution of $(4+n)$-dimensional theory. For $n=7$, the dilaton will be put 
to zero and the corresponding solution would be the SM2-brane solution of
M-theory, whereas, for $n=6$, the corresponding solution is the SD2-brane
solution of string theory. We will first show that all S2-brane solutions
in $d=n+4$ gives rise to the flat, homogeneous, isotropic 4-dimensional 
universe
of the type discussed by TW. Then we show that for $n=7,\,6$ both\footnote{
One may think that since SD2-brane solution can be constructed from SM2
solution by an analogue of direct dimensional reduction as in the static
case, it is not necessary to study both the cases together. However we point
out that there are more parameters in the SD2 solution than in the SM2 
solution. It is not clear how new parameters would be generated by dimensional
reduction and it would be interesting to understand reduction procedure
in this case.} the 
SM2-brane and SD2-brane solutions give accelerating cosmologies in Einstein
frame for only hyperbolic and flat space compactifications
with time varying volume.
We would like to point out that in the 
context of SM2-brane solution ($n=7$) accelerating cosmologies arising for both
hyperbolic and flat space compactifications were first shown by Ohta
\cite{ohta}. However, the solution used in \cite{ohta}
has a slightly different form from the Chen-Gal'tsov-Gutperle (CGG) 
\cite{cgg} solution  
we use here to demonstrate the accelerating cosmology
in the sense that the field-strength in
the former case has been chosen to a particular value which is not necessary 
in our analysis.  

The action for the $(4+n)$-dimensional gravity coupled to the dilaton and 
an $n$-form field
strength in Einstein frame is given by,
\be
S_{4+n} = \frac{1}{16\pi G_{4+n}}\int d^{4+n}x \sqrt{-g_{(4+n)}}\left(
R_{(4+n)} - \frac{1}{2} \partial_\mu \phi \partial^\mu \phi - \frac{1}
{2\cdot n!}e^{a\phi} F_n^2\right)
\ee
This action has been solved by CGG \cite{cgg} and the solution is given as, 
\bea
ds^2 &=& - e^{2A(t)} dt^2 + e^{2B(t)} dx_{(3)}^2 + e^{2C(t)} 
d\Sigma_{n,\sigma}^2\nonumber\\
\phi(t) &=& \frac{a(n+2)}{(n-1)} B(t) + c_1 t + c_2\nonumber\\
F_n &=& b\,\, \epsilon(\Sigma_{n,\sigma}) 
\eea
Here $A(t)$, $B(t)$, $C(t)$ are functions of $t$ which are chosen to satisfy
a gauge condition $-A+3B+nC=0$ to simplify the equations of motion. So, the
functions $A,\,B,\,C$ can be expressed in terms of two functions $f(t)$ and 
$g(t)$ as follows,
\be
A(t) = n g(t) - \frac{3}{(n-1)} f(t), \qquad 
B(t) = f(t), \qquad C(t) = g(t) - \frac{3}{(n-1)} f(t)
\ee
Also in the above $dx_{(3)}^2$ is the line element of 3-dimensional
Euclidean space and $d\Sigma_{n,\sigma}^2$ is that of the hyperbolic space
(for $\sigma = -1$), flat space (for $\sigma$ = 0) and spherical space
(for $\sigma = +1$), with $R_{ab} = \sigma(n-1)\bar{g}_{ab}$. 
$a$ is the dilaton coupling to the $n$-form field 
strength and $c_1$, $c_2$ are integration constants. $b$ is the field
strength parameter and $\epsilon(\Sigma_{n,\sigma})$ is the unit volume form
of $\Sigma_{n,\sigma}$.

By solving the field equations the functions $f(t)$ and $g(t)$ can be 
obtained as,
\bea
f(t) &=& \frac{2}{\chi} \ln \frac{\alpha}{\cosh[\frac{\chi\alpha}{2}
(t - t_0)]} + \frac{1}{\chi} \ln \frac{(n+2) \chi}{(n-1) b^2} - 
\frac{a}{\chi}(c_1t + c_2)\\
g(t) &=& \cases{\frac{1}{(n-1)} \ln \frac{\beta}{\sinh [(n-1) \beta |t
-t_1|]}, & {\rm for} $\quad \sigma = -1$\cr
\pm \beta(t-t_1), & {\rm for} $\quad \sigma = 0$\cr
\frac{1}{(n-1)} \ln \frac{\beta}{\cosh [(n-1) \beta (t
-t_1)]}, & {\rm for} $\quad \sigma = +1$\cr}
\eea
where $\chi=6+a^2(n+2)/(n-1)$ and the constants $\alpha$, $\beta$ and $c_1$
satisfy
\be
\frac{3c_1^2}{\chi}+\frac{(n+2)\chi\alpha^2}{2(n-1)}-n(n-1)\beta^2=0
\ee
Also note that for an $n$-form field strength the dilaton coupling $a
=2(4-n)/(n+2)$ for $n<7$ and we will put it to zero for $n=7$. Eqs.(2) -- (6)
represent the S2-brane supergravity solution in $(4+n)$-dimensions. The metric
in (2) is given in the Einstein frame.

Now we note that using eq.(3) the metric in (2) can be written as,
\be
ds^2 = e^{-ng(t)+\frac{3n}{(n-1)}f(t)} ds_E^2 + e^{2g(t) - \frac{6}{(n-1)}
f(t)}d\Sigma_{n,\sigma}^2
\ee
where,
\be
ds_E^2 = -S^6 dt^2 + S^2 dx_{(3)}^2
\ee
and the function $S(t)$ is defined as,
\be
S(t) = e^{\frac{n}{2}g(t)} e^{-\frac{(n+2)}{2(n-1)}f(t)}
\ee
Now using (7) we can reduce the action (1) to four dimensions and it has the
form \cite{gv,eg}
\be
S_4 = \frac{1}{16\pi G_4}\int d^4x \sqrt{-g}\left(R - \frac{1}{2}\partial_\mu
\phi \partial^\mu \phi - \frac{n(n+2)}{2} \partial_\mu \psi \partial^\mu
\psi - V(\phi,\psi)\right)
\ee
where,
\be
V(\phi,\psi) = \frac{b^2}{2} e^{-\frac{2(n-4)}{n+2}\phi-3n\psi} - \sigma
n(n-1) e^{-(n+2)\psi}
\ee
The radion field $\psi$ in the above can be read off from (7) as $\psi=
g(t)-3f(t)/(n-1)$. The qualitative behavior of the 4-dimensional cosmology
can be studied \cite{eg} by analyzing the potential $V(\phi,\psi)$ given in
(11). However, we will study the cosmology directly from the 4-dimensional 
metric given in (8).

Note that the 4-dimensional metric $ds_E^2$ is given in the Einstein 
frame and it takes 
the standard form of flat, homogeneous, isotropic universe with a scale factor
$S(t)$ if we define the time coordinate $\eta$ as $d\eta = S^3(t) dt$. Thus
we have shown that S2-brane solution of $(4+n)$-dimensional theory can 
be reduced to TW form. The 4-dimensional universe will be expanding if
$\frac{dS}{d\eta}>0$, which implies
\be
m(t) = \frac{n}{2} \frac{dg}{dt} - \frac{(n+2)}{2(n-1)}\frac{df}{dt}
> 0
\ee
Furthermore the expansion will be accelerating if $\frac{d^2S}{d\eta^2}>0$,
which implies,
\be
\frac{1}{\sqrt{2}}(\frac{dm(t)}{dt})^{1/2} - m(t) > 0
\ee
We will first look at the case when the dilaton is vanishing. From the 
dilaton equation in (2), we have $a=c_1=c_2=0$. Also from (6) we get for 
$\alpha = 1$\footnote{It is easy to see that for general $\alpha$, the 
arguments of the hyperbolic functions appearing in (17) and (18) will 
simply be multiplied by this parameter, but this parameter can be absorbed
by scaling $t$ appropriately.}
\be
\beta = \frac{\sqrt{3(n+2)/n}}{(n-1)}
\ee
Note that we have $\chi = 6$ in this case. So, the function $f(t)$ in 
(4) reduces to 
\be
f(t) = -\frac{1}{6}\ln \frac{b^2 (n-1) \cosh^2 3t}{6(n+2)}
\ee
In the above we have set $t_0=0$. Now in the following we consider the three
cases $\sigma=-1,0,+1$ separately\footnote{Various properties of the 
hyperbolic space compactification in the cosmological context can be found
in \cite{kmrst}.}.

\vs{2}
\noindent{\it (a) Internal space is hyperbolic ($\sigma = - 1$)}
\vs{2}

In this case the function $g(t)$ is given as
\be
g(t) = \frac{1}{(n-1)} \ln \frac{\sqrt{3(n+2)/n}}{(n-1)\sinh(\sqrt{3(n+2)/n}
|t|)}
\ee
For the accelerating expansion the conditions (12) and (13) in this case
give,
\bea
m(t) &=& - \frac{\sqrt{3n(n+2)}}{2(n-1)}\coth(\sqrt{3(n+2)/n} t) + \frac{
(n+2)}{2(n-1)} \tanh 3t > 0\\
& & \frac{\sqrt{3}}{2} \sqrt{\frac{(n+2)}{(n-1)}}\left(\frac{1}{\sinh^2
(\sqrt{3(n+2)/n}t)} + \frac{1}{\cosh^2 3t}\right)^{1/2} - m(t) > 0
\eea
For $n=7$ the above conditions reduce exactly to the conditions (17) and
(18) of ref.\cite{ohta}. However, note that in our case the metric 
as well as the
scale factor $S(t)$ in (9) depend explicitly on the field strength parameter
$b$ through the function $f(t)$ given in (15). But the scale factor in 
\cite{ohta}
does not appear to contain the field strength as it is set to a particular
value. In fact the term involving the field strength is simply an additive
constant to the function $f(t)$ and therefore does not affect the conditions
(12), (13) for the accelerating expansion. It is therefore not necessary to set
it to a particular value. Since the solution of (17)
and (18) has already been studied in \cite{ohta}, for $n=7$, 
we will not repeat it here.
We just mention that the above conditions could be satisfied simultaneously
for only negative value of $t$ in a certain interval. In terms of the true time
coordinate $\eta$ of the 4-dimensional world, this means accelerating 
expansion occurs for some interval in the positive time. The plots of 
the l.h.s.
of (17) and (18) for $n=7$ were given in \cite{ohta}.

\vs{2}
\noindent{\it (b) Internal space is flat ($\sigma=0$)}
\vs{2}

In this case $g(t) = \pm \frac{\sqrt{3(n+2)/n}}{(n-1)}t$, where we have taken
the same value of $\beta$ as in (14). Then the conditions (12) and (13) give,
\bea
m(t) &=& \pm \frac{1}{2}\frac{\sqrt{3n(n+2)}}{(n-1)} + \frac{
(n+2)}{2(n-1)} \tanh 3t > 0\\
& & \frac{\sqrt{3}}{2} \sqrt{\frac{(n+2)}{(n-1)}}\frac{1}
{\cosh 3t} - m(t) > 0
\eea
For $n=7$, these conditions are exactly the same as in \cite{ohta}. 
An accelerated
expansion has been found for a certain period of $t$ in the region $t<0$, with
the positive sign of the first term in (19). The solutions of these equations
for $n=7$ has been studied and the plot of the functions on the l.h.s. of the
conditions (19) and (20) showing an accelerated expansion has been given 
there. Unlike in the hyperbolic case there is no singularity at $t=0$ here.

\vs{2}
\noindent{\it (c) Internal space is spherical ($\sigma = +1$)}
\vs{2}

In this case the conditions (12) and (13) can not be satisfied simultaneously,
indicating that there is no accelerated expansion of the 4-dimensional universe
when the theory is compactified on spherical space\footnote{However, we would
like to point out that this conclusion is valid when the parameter $t_0=0$.
An accelerating expansion can be found even in this case for $t_0 < 0$ as
noted in \cite{eg,no}.}.

Let us now look at the case when the dilaton is non-vanishing. For $n=6$, the
corresponding solution is SD2-brane of type IIA string theory. We will find
a very similar behavior of the 4-dimensional universe as in SM2 case. The
function $f(t)$ in this case has the form,
\be
f(t) = -\frac{1}{\chi}\ln \frac{\cosh^2(\chi\alpha t/2) e^{\frac{2(4-n)}
{(n+2)}c_1t} b^2 (n-1)}{\alpha^2 (n+2) \chi}
\ee
The value of $\chi$ and dilaton coupling $a$ were given before. We have
chosen $c_2=t_0=0$\footnote{Note that when $c_2=0$, the parameter $\beta$ 
can not be removed by renaming other parameters as is done in \cite{roy}.} 
and the value 
of the parameter $\beta$ is chosen\footnote{We have chosen this value for
convenience. Here unlike in SM2 case $\beta$ does not get fixed to this 
particular value and there is a freedom. However, the essential behavior for
the conditions of accelerating expansion does not change much for different 
values of $\beta$.} as
before in (14). Then the relation between the parameters $c_1$ and $\alpha$
is given from (6) as,
\be
\frac{3 c_1^2}{\chi} + \frac{(n+2) \chi \alpha^2}{2(n-1)} - \frac{3(n+2)}
{(n-1)} = 0
\ee
Again we will discuss the three cases $\sigma = -1, 0, +1$ separately.

\vs{2}
\noindent{\it (a') Internal space is hyperbolic ($\sigma = - 1$)}
\vs{2}

The function $g(t)$ has exactly the same form as in case (a) before. The 
conditions for the accelerated expansion (12) and (13) therefore give,
\bea
&& m(t) = \nonumber\\
&& - \frac{\sqrt{3n(n+2)}}{2(n-1)}\coth(\sqrt{3(n+2)/n} t) + \frac{
(n+2)\alpha}{2(n-1)} \tanh(\chi\alpha t/2) + \frac{(4-n)}{(n-1)}\frac{c_1}
{\chi}> 0\\
&& \frac{1}{2} \sqrt{\frac{(n+2)}{(n-1)}}\left(\frac{3}{\sinh^2
(\sqrt{3(n+2)/n}t)} + \frac{\chi\alpha^2}{2}\frac{1}{\cosh^2(\chi\alpha t/2)}
\right)^{1/2} - m(t) > 0
\eea
For $n=6$, $\chi = 32/5$ and the relation (22) simplifies to
\be
\frac{15c_1^2}{32} + \frac{128 \alpha^2}{25} = \frac{24}{5}
\ee
Since the parameters are real we note from (25) that $c_1$ lies between
$-3.2$ and $+3.2$ and taking $\alpha \geq 0$, we get $0 \leq \alpha \leq 
0.97$. The conditions (23) and (24) simplify for $n=6$ to,
\bea
m(t) &=& 
 - \frac{6}{5}\coth(2t) + \frac{4
\alpha}{5} \tanh(16\alpha t/5) - \frac{c_1}
{16}> 0\\
& & \sqrt{\frac{2}{5}}\left(\frac{3}{\sinh^2
(2t)} + \frac{16\alpha^2}{5}\frac{1}{\cosh^2(16\alpha t/5)}
\right)^{1/2} - m(t) > 0
\eea
We do not get solutions of the above two conditions simultaneously for the 
whole range of parameters $c_1$ and $\alpha$ mentioned above. We plot the 
l.h.s. of both the conditions (26) and (27) versus $t$ in fig.1 for $c_1=1$
and $\alpha=0.92$. The plot shows very similar behavior as those obtained
in refs.\cite{tw,ohta} i.e. the conditions are satisfied simultaneously 
for some
interval of negative $t$ and beyond that the universe is decelerating
for both $t\to 0^-$ ($\eta \to \infty$) and $t \to -\infty$ ($\eta \to 0$).
Note that there is a singularity of the scale factor $S(t)$ at $t=0$. However,
this remains unobservable since it is at an infinite proper time in the
future of any event with $t<0$ and an infinite proper time in the past of
any event with $t>0$. So, our universe will separate into two regions with
$t<0$ and $t>0$.
Also note that we have studied the evolution for particular values
of the parameters $c_1$ and $\alpha$. There are in fact a large class
of solutions of the 4-dimensional world which would exhibit accelerated
expansions. The values of the parameters might get fixed by the detailed
knowledge of the evolution of the universe.

\begin{figure}
\begin{center}
\psfig{figure=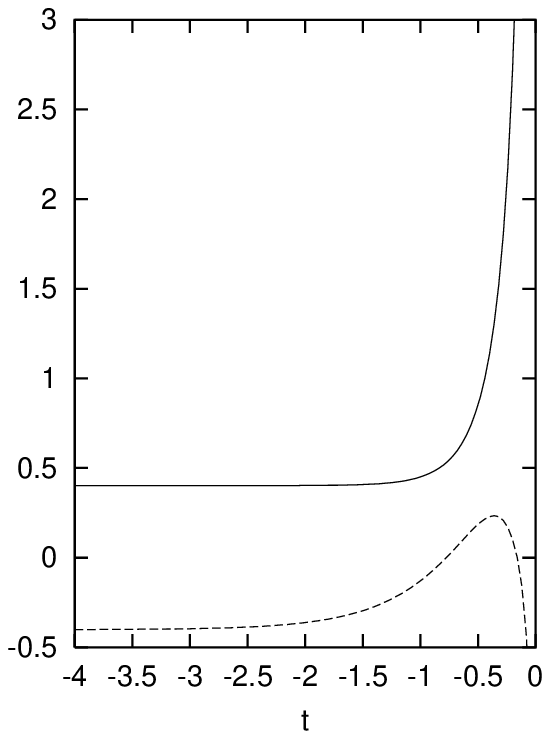,width=12cm,height=8cm}
\caption{The function $m(t)$ in eq.(26) and the l.h.s. of eq.(27) are plotted
against $t$ for $c_1=1$ and $\alpha=0.92$ and are given respectively by solid
and dashed lines.}
\end{center}
\end{figure}

\vs{2}
\noindent{\it (b') Internal space is flat ($\sigma=0$)}
\vs{2}

The function $g(t)$ in this case is as given in (b) above. The conditions (12) 
and (13) give,
\bea
m(t) &=& \pm \frac{1}{2}\frac{\sqrt{3n(n+2)}}{(n-1)} + \frac{
(n+2)\alpha}{2(n-1)} \tanh(\chi\alpha t/2) + \frac{(4-n)}{(n-1)} \frac{c_1}
{\chi} > 0\\
& & \frac{1}{2\sqrt{2}} [\frac{(n+2)\chi\alpha^2}{(n-1)}]^{1/2}\frac{1}
{\cosh(\chi\alpha t/2)} - m(t) > 0
\eea
The parameters $c_1$ and $\alpha$ are related by eq.(22). For $n=6$ we get
from above,
\bea
m(t) &=& \pm \frac{6}{5} + \frac{
4\alpha}{5} \tanh(16\alpha t/5) - \frac{c_1}
{16} > 0\\
& & \frac{4\sqrt{2}}{5}\alpha \frac{1}
{\cosh(16\alpha t/5)} - m(t) > 0
\eea
with the parameter relation given in eq.(25). Here also we do not get the 
solutions of (30) and (31) in the whole range of parameters $c_1$ and $\alpha$
given earlier. We plot the l.h.s. of both the conditions (30) and (31) in
fig.2 for $c_1=1$ and $\alpha=0.92$ and we get an accelerated expansion
only for the positive sign of the first term in (30). Here there is no 
singularity of the functions in the l.h.s. of (30) and (31) at $t=0$ unlike 
in the hyperbolic case. The accelerated expansion occurs at certain interval
of negative $t$ as is shown in fig.2. The universe decelerates at 
both $t \to -\infty$ ($\eta \to
0$) and $t \to \infty$ ($\eta \to \infty$). This range is continuously 
connected through $t=0$. A similar behavior was observed for SM2 case in
\cite{ohta}.

\begin{figure}
\begin{center}
\psfig{figure=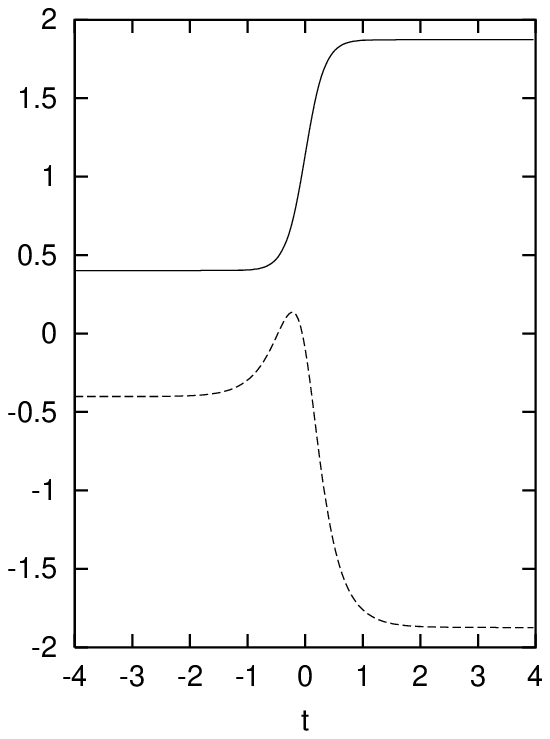,width=12cm,height=8cm}
\caption{The function $m(t)$ in eq.(30) and the l.h.s. of eq.(31) are plotted
against $t$ for $c_1=1$ and $\alpha=0.92$ and are given respectively by solid
and dashed lines.}
\end{center}
\end{figure}

\vs{2}
\noindent{\it (c') Internal space is spherical ($\sigma = +1$)}
\vs{2}

Here the function $g(t)$ is given as,
\be
g(t) = \frac{1}{(n-1)} \ln \frac{\sqrt{3(n+2)/n}}{(n-1)\cosh(\sqrt{3(n+2)/n}
t)}
\ee
where we have used the same value of $\beta$ as given in (14). So, the
conditions (12) and (13) give,
\bea
&& m(t) = \nonumber\\
&& - \frac{\sqrt{3n(n+2)}}{2(n-1)}\tanh(\sqrt{3(n+2)/n} t) + \frac{
(n+2)\alpha}{2(n-1)} \tanh(\chi\alpha t/2) + \frac{(4-n)}{(n-1)}\frac{c_1}
{\chi}> 0\\
&& \frac{1}{2} \sqrt{\frac{(n+2)}{(n-1)}}\left(-\frac{3}{\cosh^2
(\sqrt{3(n+2)/n}t)} + \frac{\chi\alpha^2}{2}\frac{1}{\cosh^2(\chi\alpha t/2)}
\right)^{1/2} - m(t) > 0
\eea
For $n=6$ they simplify to,
\bea
m(t) &=& 
 - \frac{6}{5}\tanh(2t) + \frac{4
\alpha}{5} \tanh(16\alpha t/5) - \frac{c_1}
{16}> 0\\
& & \sqrt{\frac{2}{5}}\left(-\frac{3}{\cosh^2
(2t)} + \frac{16\alpha^2}{5}\frac{1}{\cosh^2(16\alpha t/5)}
\right)^{1/2} - m(t) > 0
\eea
The parameters $c_1$ and $\alpha$ are related by eq.(25). We have checked that 
these conditions can not be satisfied simultaneously either for $t<0$ or 
for $t>0$.
So, there is no accelerated expansion of the 4-dimensional universe when the
theory is compactified on spherical spaces. Here again this conclusion is
true only for the parameter $t_0=0$. But when $t_0<0$, an accelerating 
expansion in this case has been obtained in \cite{no}.

To summarize, we have shown that $(4+n)$-dimensional gravity coupled to the
dilaton and an $n$-form field strength can give rise to a flat 4-dimensional
universe with a scale factor of the type discussed by TW under time dependent
compactification. The solutions are S2-brane solutions of $(4+n)$-dimensional
theory. For $n=7$ (when the dilaton is put to zero) the corresponding SM2-brane
solution is shown to give accelerated expansion of the 4-dimensional universe
when the internal space is both hyperbolic and flat with time varying volume. 
Similar
conclusions were found even for $n=6$ i.e. for SD2-brane case. Here there is 
a freedom of the choice of the parameters $c_1$ and $\alpha$ whose exact
values might be determined by the detailed knowledge of the evolution of our
universe. 

\vspace{.5cm}

\noindent{\bf Note added:}

\vspace{.2cm}

After submission of the paper to the net I was informed by Lorenzo 
Cornalba and Miguel Costa that observations on accelerating cosmologies
for a similar theory (studied in this paper) compactified on the time 
dependent flat internal space have been made in \cite{corcos}.  

\section*{Acknowledgements}

I would like to thank Debades Bandyopadhyay for his generous help to plot the
functions in figs.1,2. I would also like to thank Nobuyoshi Ohta and Mattias
Wohlfarth for 
very useful correspondences.


\begin{thebibliography}{99}

\bibitem{perl} S. Perlmutter et. al. ``Measurement of cosmological parameters
omega and lambda from the first 7 supernovae $Z \geq 0.35$'', Astrophys. J.
483 (1997) 565, [astro-ph/9608192]; B. Schmidt et. al. ``The high-Z supernova
search: measuring cosmic deceleration and global curvature of the universe 
using type Ia supernova'', Astrophys. J. 507 (1998) 46, [astro-ph/9805200];
A. Riess et. al. ``Observational evidence 
from supernovae for an accelerating universe and a cosmological constant'',
Astron. J. 116 (1998) 1009, [astro-ph/9805201]; P. Garnavich et. al. 
``Supernova limits on the cosmic equation of state'', Astrophys. J. 509 (1998)
74, [astro-ph/9806396]; S. Perlmutter et. al. ``Measurement of omega and lambda
from 42 high redshift supernovae'', Astrophys. J. 517 (1999) 565, 
[astro-ph/9812133].
 
\bibitem{rie} A. Riess et. al. ``The furthest known supernova: support for an 
accelerating universe and a glimpse of the epoch of deceleration'', Astrophys.
J. 560 (2001) 49, [astro-ph/0104455].

\bibitem{lb} A. Lewis and S. Bridle, ``Cosmological parameters from CMB and 
other data: a Monte-Carlo approach'', Phys. Rev. D66 (2002) 103511,
[astro-ph/0205436].

\bibitem{be} C. Bennett et. al. ``First year Wilkinson Microwave Anisotropy
Probe (WMAP) observations: preliminary maps and basic results'', 
astro-ph/0302207.

\bibitem{mn} J. Maldacena and C, Nu\~ nez, ``Supergravity description of field
theories on curved manifolds and a no-go theorem'', Int. J. Mod. Phys. A16
(2001) 822, [hep-th/0007018].

\bibitem{tow} P. Townsend, ``Quintessence from M-theory'', JHEP 11 (2001) 042,
[hep-th/0110072].

\bibitem{gibhul} G. Gibbons and C. Hull, ``de Sitter space from warped 
supergravity solutions'', hep-th/0111072.

\bibitem{tw} P. Townsend and M. Wohlfarth, ``Accelerating cosmologies from
compactification'', hep-th/0303097.

\bibitem{cgg} C. -M. Chen, D. Gal'tsov and M. Gutperle, ``S-brane solutions in 
supergravity theories'', Phys. Rev. D66 (2002) 024043, [hep-th/0204071]. 

\bibitem{gs}  M. Gutperle and A. Strominger, ``Space-like branes'', JHEP 04 
(2002) 018, [hep-th/0202221].

\bibitem{kmp} M. Kruczenski, R. Myers and A. Peet, ``Supergravity S-branes'',
JHEP 05 (2002) 039, [hep-th/0204144].

\bibitem{roy} S. Roy, ``On supergravity solutions of space-like D$p$-branes'',
JHEP 08 (2002) 025, [hep-th/0205198].

\bibitem{dk} N. Deger and A. Kaya, ``Intersecting S-brane solutions of D = 11
supergravity'', JHEP 07 (2002), [hep-th/0206057]; J. Wang, ``Spacelike and 
time dependent branes from DBI'',
JHEP 10 (2002) 037, [hep-th/0207089]; F. Quevedo, S. Rey, G. Tasinato and
I Zavala, ``Cosmological spacetimes from negative tension brane backgrounds'',
JHEP 10 (2002) 028, [hep-th/0207104];
A. Buchel, P. Langfelder, and J. Walcher, ``Does the tachyon 
matter?'', Annals Phys. 302 (2002) 78, [hep-th/0207235];
V. Ivashchuk, ``Composite S-brane solutions related to Toda type
system'', Class. Quant. Grav. 20 (2003) 261, [hep-th/0208101];
A. Strominger, ``Open string creation by S-branes'', 
hep-th/0209090; F. Quevedo, G. Tasinato and I. Zavala, ``S-branes, negative
tension branes and cosmology, hep-th/0211031;
K. Hashimoto, P. Ho and J. Wang, ``S-brane actions'', 
hep-th/0211090;
A. Buchel and J. Walcher, ``The tachyon does matter'', hep-th/
0212150;
N. Ohta, ``Intersection rules for S-branes'', hep-th/0301095;
C. Burgess, P. Martineau, F. Quevedo, G. Tasinato and I. Zavala,
``Instabilities and particle production in S-brane geometries'', 
hep-th/0301095;
A. Maloney, A. Strominger and X. Yin, ``S-brane thermodynamics'',
hep-th/0302146;
F. Leblond and A. Peet, ``SD-brane gravity fields and rolling 
tachyons'', hep-th/0303035;
K. Hashimoto, P. Ho, S. Nagaoka and J. Wang, ``Time evolution 
via S-branes'', hep-th/0303172; N. Deger, ``Non-standard intersections of 
S-branes in $D=11$ supergravity'', hep-th/0303232.

\bibitem{sen} A. Sen, ``Non-BPS states and branes in string theory'',
hep-th/9904207.

\bibitem{str} A. Strominger, ``The dS/CFT correspondence'',
JHEP 10 (2001) 034, [hep-th/0106113].

\bibitem{ohta} N. Ohta, ``Accelerating cosmologies from S-branes'',
hep-th/0303238.

\bibitem{gv} J. Garriga and A. Vilenkin, ``Solutions to the cosmological 
constant problems'', Phys. Rev D64 (2001) 023517, [hep-th/0011262].

\bibitem{eg} R. Emparan and J. Garriga, ``A note on accelerating cosmologies
from compactifications and S-branes'', hep-th/0304124.

\bibitem{kmrst} N. Kaloper, J. March-Russell, G. Starkman and M. Trodden,
``Compact hyperbolic extra dimensions: branes, Kaluza-Klein modes and
cosmology'', Phys. Rev. Lett. 85 (2000) 928, [hep-ph/0002001]; S. Nasri,
P. Silva, G. Starkman and M. Trodden, ``Radion stabilization in compact
hyperbolic extra dimensions'', Phys. Rev. D66 (2002) 045029, [hep-th/0201063];
G. Starkman, D. Stojkovic and M. Trodden, ``Homogeneity, flatness and `large'
extra dimensions'', Phys. Rev. Lett. 87 (2001) 231303, [hep-th/0106143].

\bibitem{no} N. Ohta, ``A study of accelerating cosmologies from 
superstring/M theories'', hep-th/0304172.

\bibitem{corcos} L. Cornalba and M. Costa, ``A new cosmological scenario
in string theory'', Phys. Rev. D66 (2002) 066001, [hep-th/0203031];
L. Cornalba, M. Costa and C. Kounnas, ``A resolution of the cosmological 
singularity with orientifolds'', Nucl. Phys. B637 (2002) 378, [hep-th/0204261].
\end{thebibliography}
\end{document}